\documentclass[a4paper,twoside,10pt]{article}
\usepackage{graphicx,mjyu,publaob}
\usepackage{amssymb}
\usepackage{natbib}
\usepackage{epsfig,color}

\newcommand{\be}{\begin{displaymath}}
\newcommand{\ee}{\end{displaymath}}

\def\hang{\hangindent\parindent}
\def\textindent#1{\indent\llap{#1\enspace}\ignorespaces}

\def\item{\par\hang\textindent}

\def\narrower{\advance\leftskip\parindent \advance\rightskip\parindent}

\begin{document}

\markboth
{\runn MODELING OF MOLECULAR CLOUDS WITH FORMATION OF PRESTELLAR CORES} 
{\runn SAVA DONKOV, ORLIN STANCHEV \& TODOR VELTCHEV}

{\ }

\vskip 78pt
\centerline{\naslov MODELING OF MOLECULAR CLOUDS}
\centerline{\naslov WITH FORMATION OF PRESTELLAR CORES}

{\ }

\centerline{SAVA DONKOV$^1$, ORLIN STANCHEV$^2$, TODOR VELTCHEV$^2$,}

{\ }

\centerline{\it $^1$Department of Applied Physics, Technical University}
\centerline{E-mail: savadd@tu-sofia.bg}
\centerline{\it $^2$Department of Astronomy, Faculty of Physics, University of Sofia}
\centerline{E-mail: o\_stanchev@phys.uni-sofia.bg, eirene@phys.uni-sofia.bg}

{\ }

\noindent {\smallbold Abstract.} {\abs We develop a statistical approach for description of dense 
structures (cores) in molecular clouds that might be progenitors of stars. Our basic assumptions 
are a core mass-density relationship and a power-law density distribution of these objects as 
testified by numerical simulations and observations. The core mass function (CMF) was derived 
and its slope in the high-mass regime was obtained analytically. Comparisons with observational 
CMFs in several Galactic clouds are briefly presented.} 

\section{INTRODUCTION}

Molecular clouds (MCs) are the typical regions of star formation in galaxies. Recent high-resolution observational studies in the Milky Way reveal that MCs exhibit an extremely complex, clumpy and often filamentary structure (e.g. Andr\'e et al. 2010, Mensh'chikov et al. 2010), with column and spatial densities varying by many orders of magnitude. The detected large non-thermal linewidths which scale with the size of the cloud or of its larger substructures (e.g. Larson 1981, Solomon et al. 1987, Bolatto et al. 2008) have been interpreted as indicators of the presence of supersonic turbulence. Numerous works in the last two decades have demonstrated that this supersonic turbulence is among the primary physical agents regulating the birth of stars. It creates a complex network of interacting shocks, where dense cores form at the stagnation points of convergent flows. Thus, although at large scales turbulence can support MCs against contraction, at small scales it can provoke local collapse of the emerging prestellar cores. Hence, the timescale and efficiency of a protostar formation depend strongly on the wavelength and strength of turbulent driving source (Klessen, Heitsch \& Mac Low 2000, Krumholz \& McKee 2005).
 
An important structural parameter in analytical and semi-analytical models of star formation in MCs (e.g. Padoan \& Nordlund 2002, Hennebelle \& Chabrier 2009, Veltchev, Klessen \& Clark 2011) is the probability density function ($\rho$-PDF), which gives the probability to measure a given density $\rho$ in a cloud volume $dV$. As demonstrated from many numerical simulations, its shape is approximately lognormal in isothermal, turbulent media that are not significantly affected by the self-gravity (e.g. V\'azquez-Semadeni 1994, Padoan, Nordlund \& Jones 1997, Ostriker, Gammie \& Stone 1999, Federrath, Klessen \& Schmidt 2008). Its lognormality should correspond to the same feature of the observed probability distributions of the {\it column} density $N$ ($N$-PDFs) in MCs, due to the correlation between the local values of $\rho$ along a single line of sight (V\'azquez-Semadeni \& Garc\'ia, 2001). 

On the other hand, it has been argued that the PDF displays scale-dependent features and/or its shape evolves significantly in time (Federrath, Klessen \& Schmidt 2008, Pineda et al. 2010). The lognormality is typical in the low-density, predominantly turbulent regime, whereas at higher column densities a power-law tail is emerging. Such high-density power-law (PL) tail is a characteristic feature of $N$-PDFs in evolved MCs where star formation processes already occur or just start (Kainulainen et al. 2009, Froebrich \& Rowles 2010). That is confirmed as well from analysis of numerical simulations of clouds dominated by gravity (Ballesteros-Paredes et al. 2011). A consistent theory of cloud structure must take into account the characteristics sketched above. Probing the $N$- and $\rho$-PDFs can be used to set up constraints to analytical star formation theories.

In this work a statistical approach is suggested to derive the mass function of high-density (prestellar) cores that are possible progenitors of stars. Our starting point is a description of the PL tail of the $\rho$-PDF that is representative for the MC regions populated by these objects.

\section{STATISTICAL DESCRIPTION OF DENSE CORES}

\subsection{Description of the high-density power-law tail of the PDF}

Kritsuk, Norman \& Wagner (2011) showed that the $\rho$-PDF evolves from purely lognormal shape to a combination of a lognormal `hat' and a PL tail. Schematic representation of this distribution is given in Fig. 1. We use a standard designation of the logarithmic normalized density: $s\equiv\lg(\rho/\rho_0)$, where $\rho_0$ is a normalization unit. Parameters of the PL tail are the lower ($s_{\rm low}\equiv \lg(\rho_{1}/\rho_{0})$) and the upper limit ($s_{\rm up}\equiv \lg(\rho_{2}/\rho_{0})$) of the high-density range and its slope $q$. Thus, the PL tail is described by:
\begin{equation}
\label{eq_PL_tail}
 dP_s=A_s\,10^{qs}\,ds=A_s\Big(\frac{\rho}{\rho_0}\Big)^q\,d\lg\Big(\frac{\rho}{\rho_0}\Big)~,
\end{equation}
where the coefficient $A_s=(q\ln10)/(10^{qs_{\rm up}}-10^{qs_{\rm low}})$ is obtained from the requirement $\int_{s_{\rm low}}^{s_{\rm up}}dP_s=1$.

\vfill\eject

\centerline{\includegraphics[angle=-90, width=0.9\textwidth]{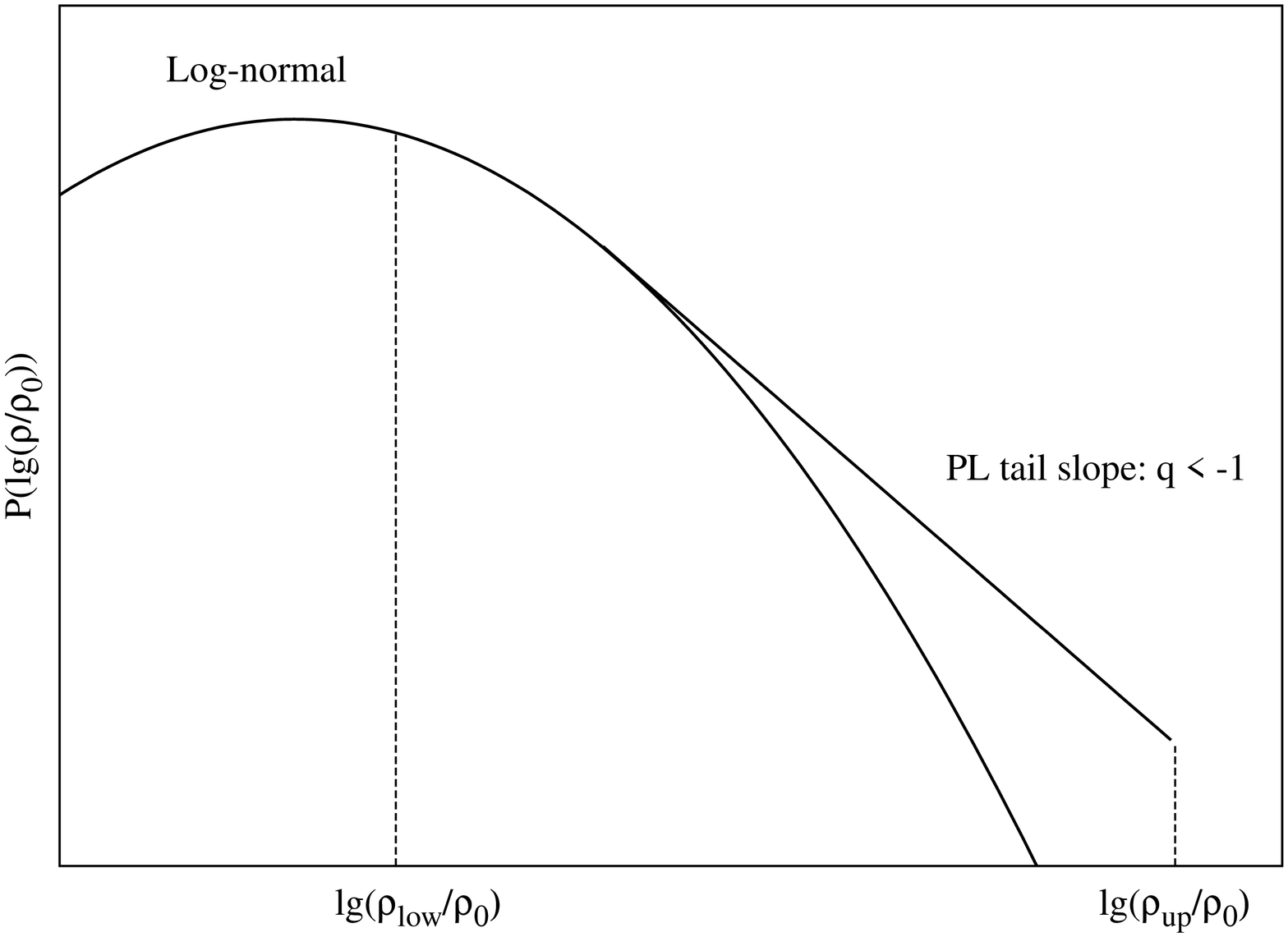}}
\vskip2mm
\noindent {\bf Figure 1.} Schematic representation of the $\rho$--PDF as a combination  of a lognormal function with a power-law tail in the high-density part. 

\subsection{Core mass-density relationship}

In observational studies, the internal structure of MCs is usually described by a sample of discrete condensations, 
delineated through different procedures and labelled `clumps' or `cores'. In our approach, the set of individual clumps 
in a considered volume is represented by an ensemble of statistical, not individual objects as the physical characteristics 
of both groups obey the same statistical relations (e.g. size-mass relation, density distribution). Then the $\rho$-PDF corresponds 
to the density distribution of statistical clumps. Those with densities in the range of the PL tail are dense cores and hereafter we 
label them just ``cores'' for simplicity.

Our basic physical assumption is the existence of a core mass-density relationship: 

\begin{equation}
\label{eq_rho-m}
\lg\Big(\frac{\rho}{\rho_0}\Big)=x\,\lg\Big(\frac{m}{m_0}\Big)
\end{equation}
where $m_0$ is an arbitrary unit of normalization and the power-law index $x$ is {\it negative} and assumed to be fixed within 
the whole PL tail (Lombardi, Alves \& Lada, 201). Further we adopt the natural presupposition about a statistical relation 
between core masses $m$, densities $\rho$ and sizes $l$,
\begin{equation}
 \label{eq_m-rho-l}
 \frac{m}{m_0}=\frac{\rho}{\rho_0} \cdot \,\Big(\frac{l}{l_0}\Big)^3~,
\end{equation}
and obtain by use of Eq.~\ref{eq_rho-m} core size-density and size-mass (with size normalization unit $l_0$) as well:
\begin{equation}
\label{eq_rho-l}
\lg\Big(\frac{\rho}{\rho_0}\Big)=\frac{3x}{1-x}\,\lg\Big(\frac{l}{l_0}\Big)~,
\end{equation}
\begin{equation}
\label{eq_m-l}
\lg\Big(\frac{m}{m_0}\Big)=\frac{3}{1-x}\,\lg\Big(\frac{l}{l_0}\Big)~.
\end{equation}

Taking into account the one-to-one correspondence between the core quantities $\rho$, $m$ and $l$, one derives from Eqs. \ref{eq_PL_tail}, \ref{eq_rho-m} and \ref{eq_rho-l} the probability distributions of core masses and sizes:
\begin{equation}
 \label{eq_PL_m_tail}
 dP(m)=A_m\,\Big(\frac{m}{m_0}\Big)^{qx}\,d\lg\Big(\frac{m}{m_0}\Big)~,
\end{equation}
\begin{equation}
 \label{eq_PL_l_tail}
 dP(l)=A_l\,\Big(\frac{l}{l_0}\Big)^{q\frac{3x}{1-x}}\,d\lg\Big(\frac{l}{l_0}\Big)~,
\end{equation}
where $A_m$ and $A_l$ are calculated from the normalization of probability measures. Note as well the obvious relation: $dP_s=dP(m)=dP(l)$. It defines the distributions of core masses and sizes in the PL tail which we label hereafter $m$-PDF and $l$-PDF, respectively.

It seems natural to choose the mean density of the cloud $\langle\rho\rangle$ and a fixed fraction $\kappa$ of its size $L$ as 
normalization units (Veltchev, Klessen \& Clark, 2011):
\begin{equation}
 \label{eq_norm_units}
 \rho_0\equiv\langle\rho\rangle~,~~~~l_0\equiv\kappa L  
\end{equation}

\subsection{Averaged core quantities}

The average core density could be calculated in two alternative ways:

{\it Arithmetic average}:
    {\begin{equation}
    \label{eq_arithmetic_average}
    \overline{\Big(\frac{\rho}{\rho_0}\Big)}_{\rm ar}\equiv A_s\,\int\limits_{\rho_1}^{\rho_2}\Big(\frac{\rho}{\rho_0}\Big)\cdot\Big(\frac{\rho}{\rho_0}\Big)^q\,d\lg\Big(\frac{\rho}{\rho_0}\Big)=\Big(\frac{q}{q+1}\Big)\Big(\frac{\rho_1}{\rho_0}\Big)\Bigg[ \frac{(\rho_2/\rho_1)^{q+1}-1}{(\rho_2/\rho_1)^q~-1}\Bigg] ,
    \end{equation}}
{\it Logarithmic average}:
it is a generalization of the geometric average in case of continuous density distribution and is defined as:
    \[
    \overline{\Big(\frac{\rho}{\rho_0}\Big)}_{\rm lg} \equiv 10^{\,\overline{\lg (\rho/\rho_0)}}~,~~~{\rm where}~~~\overline{\lg\Big(\frac{\rho}{\rho_0}\Big)}= A_s\,\int\limits_{\rho_1}^{\rho_2}\lg\Big(\frac{\rho}{\rho_0}\Big)\cdot\Big(\frac{\rho}{\rho_0}\Big)^q\,d\lg\Big(\frac{\rho}{\rho_0}\Big)~~~\Longrightarrow
    \]

\begin{equation}
 \label{eq_logarithmic_average}
 \overline{\Big(\frac{\rho}{\rho_0}\Big)}_{\rm lg}= \exp\Big(-\frac{1}{q}\Big)\cdot\Big(\frac{\rho_2}{\rho_0}\Big)^\frac{(\rho_2/\rho_0)^q}{(\rho_2/\rho_0)^q-(\rho_1/\rho_0)^q}\,\Big(\frac{\rho_1}{\rho_0}\Big)^{-\frac{(\rho_1/\rho_0)^q}{(\rho_2/\rho_0)^q-(\rho_1/\rho_0)^q}}~.
\end{equation}
Analogically, one can define and calculate arithmetic and logarithmic average of core mass ans size by use of Eqs.~\ref{eq_PL_m_tail} and \ref{eq_PL_l_tail}. It is important to note that the approach of logarithmic averaging leads to a physically natural relationship:
\begin{equation}
 \label{eq_average_mass_density_size}
 \overline{\Big(\frac{m}{m_0}\Big)}_{\rm lg}=\overline{\Big(\frac{\rho}{\rho_0}\Big)}_{\rm lg} \cdot \overline{\Big(\frac{l}{l_0}\Big)}^{~3}_{\rm lg}~.
\end{equation}
The transformation from dimensionless to physical quantities is independent on the averaging approach: $\overline{\rho}=\rho_0\,\overline{(\rho/\rho_0)}$, $\overline{m}=m_0\,\overline{(m/m_0)}$, $\overline{l}=l_0\,\overline{(l/l_0)}$. Recalling Eq.~\ref{eq_m-rho-l}, one obtains:
\begin{equation}
 \label{eq_mass_density_size}
 \frac{m}{\rho\,l^3}=\frac{m_0}{\rho_0\,l_0^3}=\frac{\overline{m}}{\bar{\rho}\,\,\overline{l}^3}~.
\end{equation}
We point out that the second sign of equality in this equation holds only in case of logarithmic averaging (cf. Eq. \ref{eq_average_mass_density_size}).

\subsection{Relation between the normalization units and the formula for total number of cores}

The core mass is a fundamental non-observable quantity and it is crucial to avoid arbitrariness of choice of its normalization unit. Therefore we derive a relation between $\rho_0$, $l_0$ and $m_0$ using the requirements for volume and mass conservation considering the whole PL tail. Its total volume $V_{\rm tot}$ is calculated as follows:
\[  V_{\rm tot}=\sum\limits_{l-PDF}l^3\,N_l=l_0^3 N_{\rm tot} \sum\limits_{l-PDF}\Big( \frac{l}{l_0}\Big)^3 \frac{N_l}{N_{\rm tot}}=l_0^3 N_{\rm tot} \int\limits_{l_1}^{l_2}\Big( \frac{l}{l_0}\Big)^3 dP(l)~, \]
where $N_l$ is the number of cores with size $l$ and $N_{\rm tot}$ is a measure of the total number of cores. The limits $l_1$ and $l_2$ correspond to the density limits $\rho_1$ and $\rho_2$ according to Eq.~\ref{eq_rho-l} and one obtains after integration:

\begin{equation}
 \label{eq_V_tot}
 V_{\rm tot}=l_0^3 N_{\rm tot}\cdot\frac{\frac{qx}{1-x}}{\frac{qx}{1-x}+1}\,\Bigg[ \frac{(\rho_2/\rho_0)^{q+\frac{1-x}{x}}-(\rho_1/\rho_0)^{q+\frac{1-x}{x}}}{(\rho_2/\rho_0)^q-(\rho_1/\rho_0)^q} \Bigg] \equiv l_0^3 N_{\rm tot}\cdot Q(q,x)~.
\end{equation}
Analogically, the total mass of the cores is calculated through:
\begin{equation}
 \label{eq_M_tot}
 M_{\rm tot}=\sum\limits_{m-PDF}m\,N_m=...=m_0 N_{\rm tot}\cdot\frac{qx}{qx+1}\,\Bigg[ \frac{(\rho_2/\rho_0)^{q+\frac{1}{x}}-(\rho_1/\rho_0)^{q+\frac{1}{x}}}{(\rho_2/\rho_0)^q-(\rho_1/\rho_0)^q} \Bigg].
\end{equation}
In view of $M_{\rm tot}=\bar{\rho}\cdot V_{\rm tot}=\rho_0\,\cdot\overline{(\rho/\rho_0)}\cdot V_{\rm tot}$, we get the relation between the normalization units:
\begin{equation}
 \label{eq_ratio_norm_units}
  \frac{m_0}{\rho_0\,l_0^3}=\frac{\rho_0}{\rho_1}\overline{\Big(\frac{\rho}{\rho_0}\Big)}\,\frac{qx+1}{(q-1)x+1}\cdot\,\Bigg[ \frac{(\rho_2/\rho_1)^{q+\frac{1-x}{x}}-1}{(\rho_2/\rho_1)^{q+\frac{1}{x}}-1} \Bigg]
\end{equation}
Referring to Eq. ~\ref{eq_V_tot} and the relation: $r \rho_{0} V = r M = M_{\rm tot} = \rho_{0}\cdot \overline{(\rho/\rho_{0})}\cdot V_{\rm tot}$ , where $M$ and $V$ are the mass and volume of the whole cloud, we obtain the formula for total core number as follows: 
\begin{equation}
 \label{eq_N_tot}
 N_{\rm tot}=\frac{r}{\kappa^3}\Bigg[ \overline{\Big(\frac{\rho}{\rho_0}\Big)} Q(q,x)\Bigg]^{-1}~,
\end{equation}
where the filling factor $r$ accounts for the mass fraction of cores within the whole cloud and the density average could be arithmetic as well logarithmic.

\section{DERIVATION OF THE CORE MASS FUNCTION}

\subsection{Total number of cores and scales defined by chosen density threshold}

We consider an MC as a hierarchical set of spatial scales. They are defined as effective sizes of subregions (or a set of subregions) in the cloud delineated through chosen density thresholds. Thus, a fixed density $\rho^\prime$ corresponds to a scale which contains the mass of the whole PL-tail substructure with densities in the range $\rho^\prime\le\rho\le\rho_2$. Since the mass-density power-law index $x$ is negative, the large the density of a core the lower its mass. To derive the cumulative CMF correctly, one must take into account the contribution of each spatial scale in the hierarchical structure defined by threshold $\rho^{\prime}$. The latter is a product of the total number of scales $N_{scales}(\rho^{\prime})$ contained in the volume over the threshold $\rho^{\prime}$ and the total number of cores $N_{\rm tot}(\rho^{\prime})$ at each scale. Repeating the procedure in the previous sections for a subsection of the PL tail $[\rho^\prime,\rho_2]$, one obtains $N_{\rm tot}(\rho^\prime)$ according to Eq.~\ref{eq_N_tot} while $r$, $\overline{(\rho/\rho_0)}$ and $Q(q,x)$ are functions of the given $\rho^\prime$ (instead of $\rho_1$):
\begin{equation}
 \label{eq_N_tot_rhop}
 N_{\rm tot}(\rho^\prime)=\frac{r(\rho^\prime)}{\kappa^3}\Bigg[ \overline{\Big(\frac{\rho}{\rho_0}\Big)}(\rho^\prime)\cdot Q(q,x,\rho^\prime)\Bigg]^{-1}~.
\end{equation}
Letting $\rho^\prime\ll\rho_2$, it follows in both approaches of averaging that $\overline{(\rho/\rho_0)}(\rho^\prime)\propto(\rho^\prime/\rho_0)$ (Eqs.~\ref{eq_arithmetic_average} and \ref{eq_logarithmic_average}) and $Q(q,x,\rho^\prime)\propto(\rho^\prime/\rho_0)^{(1-x)/x}$ (Eq.~\ref{eq_V_tot}; in this context it is important to mention that simulations predict $q\leq -1.5$). In this case $N_{\rm tot}(\rho^\prime)\propto (r(\rho^\prime)/\kappa^3)\cdot(\rho^\prime/\rho_0)^{-1/x}$.

The total number of scales $N_{\rm scales}(\rho^\prime)$ is easily calculated from the filling factor $r$ and the requirement of mass conservation at each density threshold $\rho^\prime$. The total mass of cores over the given threshold $\rho^\prime$ in MC with mass $M$ and volume $V$ is $\rho_0\cdot\overline{(\rho/\rho_0)}(\rho^\prime)\cdot V_{\rm tot}(\rho^{\prime})=M_{\rm tot}(\rho^\prime)=r(\rho^\prime)M=r(\rho^\prime)\rho_0 V$ whence 
\[ V_{\rm tot}(\rho^\prime)=r(\rho^\prime)\cdot \Bigg[\overline{\Big(\frac{\rho}{\rho_0}\Big)}(\rho^\prime)\Bigg]^{-1}V \]

Obviously $r\rho_0 V=M_{\rm tot}=N_{\rm scales}(\rho^\prime)\cdot M_{\rm tot}(\rho^\prime)=N_{scales}(\rho^{\prime})\cdot \rho_{0}\cdot \overline{(\rho/\rho_{0})}(\rho^{\prime})\cdot V_{\rm tot}(\rho^{\prime})$ and after simple transformations one gets finally:
\begin{equation}
 \label{eq_N_scales_rhop}
  N_{\rm scales}(\rho^\prime)=\frac{rV}{\overline{\Big(\frac{\rho}{\rho_0}\Big)}(\rho^\prime)\cdot V_{\rm tot}(\rho^\prime)}=...=\frac{r}{r(\rho^\prime)}
\end{equation}
\newpage

\subsection{The cumulative CMF}

The cumulative CMF ${\cal N}(m^\prime)$ is derived by counting all cores with masses in the range $m^\prime \ge m \ge m_2$, corresponding to densities $\rho^\prime \le \rho \le \rho_2$. In our case, one has to multiply the total numbers of cores and scales over threshold $\rho^\prime=\rho_0(m^\prime/m_0)^x$ (Eqs. \ref{eq_N_tot_rhop} and \ref{eq_N_scales_rhop}):
\begin{eqnarray}
 \label{eq_cumul_CMF}
   {\cal N}(m^\prime) = \frac{r}{\kappa^3}\Bigg[ \overline{\Big(\frac{\rho}{\rho_0}\Big)}(\rho^\prime)\cdot Q(q,x,\rho^\prime)\Bigg]^{-1}  & \propto & \frac{r}{\kappa^3}\Big(\frac{\rho^\prime}{\rho_0}\Big)^{-\frac{1}{x}}  =  \frac{r}{\kappa^3}\Big(\frac{m^\prime}{m_0}\Big)^{-1} \\
   ~ & (\rho^\prime\ll\rho_2) & ~ \nonumber 
\end{eqnarray}

Note that the slope $\Gamma$ of a power-law differential CMF is exactly the same of its corresponding cumulative function. In other words, we derived a differential CMF with slope $-1$, typical for fractal structures (Elmegreen 1997). 

In view of the high densities in the PL tail, we may assume that most (if not all) cores are gravitationally unstable and contract in timescales given by the free-fall time $\tau_{\rm ff}\propto \rho^{-1/2}$. Then a time-weighted CMF would be more representative for results one can expect from observations. Such time weighting can be done introducing a weighting factor of each scale proportional to $\tau_{\rm ff}^{-1}\propto \overline{(\rho^\prime/\rho_0)}^{1/2}\propto (m^\prime/m_0)^{x/2}$. Then the time-weighted cumulative CMF will have a slope modified by addend $x/2$:
\begin{eqnarray}
 \label{eq_cumul_tCMF}
   {\cal N}_{\tau}(m^\prime) & \propto & \frac{r}{\kappa^3}\Big(\frac{\rho^\prime}{\rho_0}\Big)^{-\frac{1}{x}+\frac{1}{2}}  =  \frac{r}{\kappa^3}\Big(\frac{m^\prime}{m_0}\Big)^{-1+\frac{x}{2}} \\
   ~ & (\rho^\prime\ll\rho_2) & ~ \nonumber 
\end{eqnarray}

\section{DISCUSSION}

Most recent studies dedicated to the CMF show that its high-mass slope is close or identical to that of the initial stellar mass function $\Gamma=-1.35$ (Salpeter 1955). In Table ~\ref{tab_cmf_slopes} we illustrate the predictive capability of our model in comparison with CMFs, derived in some observational works and from simulations. The derived slopes and their variations could be explained by variations of the core mass-density power-law index in
the range $0\gtrsim x \gtrsim -0.7$ (with a single exception with an extreme value) which is consistent with the typical values of this quantity for the inner parts of several MCs as derived by Donkov, Veltchev \& Klessen (2011) and testified by comparison with the observational study of Lombardi, Alves \& Lada (2010).

These preliminary results are stimulating to develop further our model, including results from recent numerical simulations and/or analytical estimates of the time evolution of the $\rho$-PDF (Girichidis et al. 2012).

\begin{table}
\caption{Slopes of CMFs derived from various authors in comparison with our model predictions.}
\smallskip
\label{tab_cmf_slopes} 
\begin{tabular}{@{}ll@{\hspace{0pt}}clc}
\hline 
\hline 
Galactic MC & Ref. & Note & Slope of the CMF & $x$ \\ 
\hline 
Pipe &	1 & ~ & $-1.35$ & $-0.7$ \vspace*{4pt}\\
Orion	& 2 & ~ & $-1.35$ & $-0.7$ \vspace*{4pt}\\
Orion A	& 3 & ~ & $-1.3 \pm 0.3$ & $-0.6\pm-0.6$ \vspace*{4pt}\\
Perseus & 4 & {\small (a) lognormal~\,~} & $-1. \pm 0.1$  & $\sim 0$ \\
~	& ~ & {\small ~~~(b) time-weighted} & $-2.15 \pm 0.08$ & $\sim -2.3$ \vspace*{4pt}\\
Ophiuchus & 5 & {\small time-weighted} & $-1.35$ & $-0.7$ \vspace*{4pt}\\
Perseus, Serpens, & 6 & ~ & $-1.3\pm0.4$ & $-0.6 \pm 0.8$ \\
Ophiuchus & ~ & ~ & ~ \vspace*{4pt}\\
Simulations & 7 & ~ & $-1.15\le \Gamma \le-1.35$ &   $-0.3\ge x\ge-0.7$ \\
(PP, PPV) & ~ & ~ & ~ & ~ \\
\hline 
\hline 
\end{tabular} 
\smallskip 
{\small [1]~Alves, Lombardi \& Lada 2007, [2]~Nutter \& Ward-Thompson 2007, [3] Ikeda \& Kitamura 2009, [4] Curtis \& Richer 2010, [5] Andr\'e et al. 2007, [6] Enoch et al. 2008, [7] Smith, Clark \& Bonnell 2008}
\smallskip 
\end{table}

{\it Acknowledgement:} T.V. acknowledges partial support by the {\em Deutsche Forschungsgemeinschaft} (DFG) under grant KL 1358/15-1. 
\\

\newpage
\vskip-.5cm


\references

Alves, J., Lombardi, M., \& Lada, C.: 2007,
The mass function of dense molecular cores and the origin of the IMF,
\journal{A\&A}, \vol{462}, L17.

Andr\'e, P., Belloche, A., Motte, F., \& Peretto, N.: 2007,
The initial conditions of star formation in the Ophiuchus main cloud: Kinematics of the protocluster condensations
\journal{A\&A}, \vol{472}, 519.

Andr\'e, Ph., Men'shchikov, A., Bontemps, S., et al.: 2010,
From filamentary clouds to prestellar cores to the stellar IMF: Initial highlights from the {\sc Herschel} Gould Belt Survey,
\journal{A\&A}, \vol{518}, L102	

Ballesteros-Paredes, J., V\'azquez-Semadeni, E., Gazol, A., Hartmann, L., Heitsch, F., Col\'in, P.: 2011,
Gravity or turbulence? - II. Evolving column density probability distribution functions in molecular clouds,
\journal{MNRAS}, \vol{416}, 1436.

Bolatto, A. D., Leroy, A. K., Rosolowsky, E., Walter, F., \& Blitz, L.: 2008,
The Resolved Properties of Extragalactic Giant Molecular Clouds,
\journal{ApJ}, \vol{868}, 9.

Curtis, E., \& Richer, J.: 2010,
The properties of SCUBA cores in the Perseus molecular cloud: the bias of clump-finding algorithms,
\journal{MNRAS}, \vol{402}, 603.

Donkov, S., Veltchev, T., \& Klessen, R. S.: 2011,
Mass-density relationship in molecular cloud clumps,
\journal{MNRAS}, \vol{418}, 916.

Elmegreen, B. G.: 1997, 
The Initial Stellar Mass Function from Random Sampling in a Turbulent Fractal Cloud
\journal{ApJ}, \vol{486}, 944.

Enoch, M., Evans, N., Sargent, A., Glenn, J., Rosolowsky, E., Myers, P.: 2008,
The Mass Distribution and Lifetime of Prestellar Cores in Perseus, Serpens, and Ophiuchus,
\journal{ApJ}, \vol{684}, 1240.

Federrath, C., Klessen, R. S., \& Schmidt, W.: 2008,
The Density Probability Distribution in Compressible Isothermal Turbulence: Solenoidal versus Compressive Forcing,
\journal{ApJ}, \vol{688}, L79.

Froebrich, D., \& Rowles, J.: 2010,
The structure of molecular clouds - II. Column density and mass distributions,
\journal{MNRAS}, \vol{406}, 1350.

Girichidis, P., Konstandin, L., Klessen, R. S., \& Whitworth, A.: 2012 (in preparation)

Goodman, A. A., Pineda, J. E., \& Schnee, S. L.: 2009, 
The "True" Column Density Distribution in Star-Forming Molecular Clouds,
\journal{ApJ}, \vol{692}, 91.

Hennebelle, P., \& Chabrier, G.: 2009,
Analytical Theory for the Initial Mass Function. II. Properties of the Flow,
\journal{ApJ}, \vol{702}, 1428.

Ikeda, N., \& Kitamura, Y.: 2009, 
A C$^{18}$O Study of the Origin of the Power-Law Nature in the Initial Mass Function,
\journal{ApJ}, \vol{705}, L95

Kainulainen, J., Beuther, H., Henning, T., Plume, R.: 2009,
Probing the evolution of molecular cloud structure. From quiescence to birth,
\journal{A\&A}, \vol{508}, L35.

Klessen, R. S., Heitsch, F., \& Mac Low, M.: 2000,
Gravitational Collapse in Turbulent Molecular Clouds. I. Gasdynamical Turbulence,
\journal{ApJ}, \vol{535}, 887.

Kritsuk, A., Norman, M., \& Wagner, R.: 2011,
On the Density Distribution in Star-forming Interstellar Clouds,
\journal{ApJ}, \vol{727}, 20.

Krumholz, M., \& McKee, C. F.: 2005, 
A General Theory of Turbulence-regulated Star Formation, from Spirals to Ultraluminous Infrared Galaxies,
\journal{ApJ}, \vol{630}, 250.

Larson, R.: 1981,
Turbulence and star formation in molecular clouds,
\journal{MNRAS}, \vol{194}, 809.

Lombardi, M., Alves, J., \& Lada, C.: 2010,
Larson's third law and the universality of molecular cloud structure,
\journal{A\&A}, \vol{519}, 7.

Men'shchikov, A., Andr\'e, Ph., Didelon, P., et al.: 2010,
Filamentary structures and compact objects in the Aquila and Polaris clouds observed by {\sc Herschel},
\journal{A\&A}, \vol{518}, L103	

Nutter, D., Ward-Thompson, D.: 2007,
A SCUBA survey of Orion - the low-mass end of the core mass function,
\journal{MNRAS}, \vol{374}, 1413.

Ostriker, E., Gammie, C., \& Stone, J.: 1999,
Kinetic and Structural Evolution of Self-gravitating, Magnetized Clouds: 2.5-dimensional Simulations of Decaying Turbulence,
\journal{ApJ}, \vol{513}, 259.

Padoan, P., Nordlund, \AA., \& Jones, B.: 1997, 
The universality of the stellar initial mass function,
\journal{MNRAS}, \vol{288}, 145.

Padoan, P., \& Nordlund, \AA: 2002,
The Stellar Initial Mass Function from Turbulent Fragmentation,
\journal{ApJ}, \vol{576}, 870.

Pineda, J., Goldsmith, P., Chapman, N., et al.: 2010, 
The Relation Between Gas and Dust in the Taurus Molecular Cloud,
\journal{ApJ}, \vol{721}, 686.

Salpeter, E.: 1955,
The Luminosity Function and Stellar Evolution,
\journal{ApJ}, \vol{121}, 161. 

Smith, R., Clark, P., \& Bonnell, I.: 2008,
The structure of molecular clouds and the universality of the clump mass function,
\journal{MNRAS}, \vol{391}, 1091. 

Solomon, P. M., Rivolo, A. R., Barrett, J., \& Yahil, A.: 1987,
Mass, luminosity, and line width relations of Galactic molecular clouds,
\journal{ApJ}, \vol{319}, 730. 

V\'azquez-Semadeni, E.: 1994,
Hierarchical Structure in Nearly Pressureless Flows as a Consequence of Self-similar Statistics, 
\journal{ApJ}, \vol{423}, 681. 

V\'azquez-Semadeni, E., \& Garc\'ia, N.: 2001,
The Probability Distribution Function of Column Density in Molecular Clouds,
\journal{ApJ}, \vol{557}, 727. 

Veltchev, T., Klessen, R., \& Clark, P.: 2011,
Stellar and substellar initial mass function: a model that implements gravoturbulent fragmentation and accretion,
\journal{MNRAS}, \vol{411}, 301. 

\endreferences


\end{document}